\documentclass[aps,prb,twocolumn,superscriptaddress,showpacs]{revtex4-1}

\usepackage{eurosym}
\usepackage{amsfonts}
\usepackage{amssymb}
\usepackage{amsmath}
\usepackage{graphicx}
\begin{document}

\title{Spin-orbit coupling enhanced superconductivity in Bi-rich compounds ABi$_{3}$ (A=Sr and Ba)}

\author{D. F. Shao}
\email[The authors contributed equally to this work.]{} 
\affiliation{Key Laboratory of Materials Physics, Institute of Solid State Physics,
Chinese Academy of Sciences, Hefei 230031, People's Republic of China}

\author{X. Luo}
\email[The authors contributed equally to this work.]{}
\affiliation{Key Laboratory of Materials Physics, Institute of Solid State Physics,
Chinese Academy of Sciences, Hefei 230031, People's Republic of China}

\author{W. J. Lu}

\email[Corresponding author:]{wjlu@issp.ac.cn}
\affiliation{Key Laboratory of Materials Physics, Institute of Solid State Physics,
Chinese Academy of Sciences, Hefei 230031, People's Republic of China}

\author{L. Hu}
\affiliation{Key Laboratory of Materials Physics, Institute of Solid State Physics,
Chinese Academy of Sciences, Hefei 230031, People's Republic of China}

\author{ X. D. Zhu}
\affiliation{High Magnetic Field Laboratory, Chinese Academy of Sciences, Hefei 230031, People's Republic of China}

\author{W. H. Song}
\affiliation{Key Laboratory of Materials Physics, Institute of Solid State Physics,
Chinese Academy of Sciences, Hefei 230031, People's Republic of China}

\author{X. B. Zhu}
\affiliation{Key Laboratory of Materials Physics, Institute of Solid State Physics,
Chinese Academy of Sciences, Hefei 230031, People's Republic of China}

\author{Y. P. Sun}

\email[Corresponding author:]{ypsun@issp.ac.cn}

\affiliation{High Magnetic Field Laboratory, Chinese Academy of Sciences, Hefei
230031, People's Republic of China}

\affiliation{Key Laboratory of Materials Physics, Institute of Solid State Physics,
Chinese Academy of Sciences, Hefei 230031, People's Republic of China}

\affiliation{Collaborative Innovation Center of Microstructures, Nanjing University, Nanjing 210093, China }

\begin{abstract}
Recently, Bi-based compounds have attracted attentions because of
the strong spin-orbit coupling (SOC). In this work, we figured out
the role of SOC in ABi$_{3}$ (A=Sr and Ba) by theoretical investigation
of the band structures, phonon properties, and electron-phonon coupling.
Without SOC, strong Fermi surface nesting leads to phonon instabilities
in ABi$_{3}$. SOC suppresses the nesting and stabilizes the structure.
Moreover, without SOC the calculation largely underestimates the superconducting
transition temperatures ($T_{c}$), while with SOC the calculated
$T_{c}$ are very close to those determined by measurements on single
crystal samples. The SOC enhanced superconductivity in ABi$_{3}$
is due to not only the SOC induced phonon softening, but also the
SOC related increase of electron-phonon coupling matrix elements.
ABi$_{3}$ can be potential platforms to construct heterostructure
of superconductor/topological insulator to realize topological superconductivity.

\end{abstract}

\maketitle
\section{Introduction}

Recently, materials with strong spin-orbit coupling (SOC) effect have
attracted a great deal of attention due to the resulted novel topological
phases. Among those materials, the heaviest group V semimetal Bi-based
compounds are mostly investigated \cite{isaeva-pssrrl}. Bi$_{2}$X$_{3}$ (X=Se,
Te) \cite{Xia-Bi2Se3, chen-Bi2Te3} and ultrathin Bi(111) Films \cite{Bi111-1,
Bi111-2, Bi111-3} are suggested to be topological insulators. Introducing
superconductivity into the topological insulator can make the topological
superconductor \cite{hasan-review,Qi-review}. The Majorana fermion
is predicted to emerge in topological superconductor, which will deepen
our understanding of quantum states of matter in physics and foster
innovations in future quantum technologies \cite{hasan-review,Qi-review,
Sasaki-review}. In principle, the topological superconductivity
can show up in doped topological insulators or at the interfaces in
a device composed by superconductor and topological insulator \cite{hasan-review,Qi-review}.
However, there are only a few systems are reported to be the promising
candidates \cite{Sasaki-review}. Doping can introduce superconductivity,
making Cu$_{x}$Bi$_{2}$Se$_{3}$ \cite{CuxBi2Se3}, Sn$_{1-x}$In$_{x}$Te
\cite{Sn-InTe}, (Pb$_{0.5}$Sn$_{0.5}$)$_{1-x}$In$_{x}$Te \cite{PbSnTe}
and Cu$_{x}$(PbSe)$_{5}$(Bi$_{2}$Se$_{3}$)$_{6}$ \cite{PbSe-Bi2Se3} potential platforms to realize
topological superconductivity \cite{Sasaki-review}. Very recently a
2D helical topological superconductor was reported to be realized
in a heterostructure sample constituting of a Bi$_{2}$Se$_{3}$ film and a s-wave
superconductor NbSe$_{2}$ \cite{Hasan-Bi2Se3-NbSe2}. More platforms still
need to be explored. Since most reported candidates of topological
superconductor are Bi-based compounds \cite{Sasaki-review}, investigating
other Bi-based superconductors is necessary.

\begin{figure}
\includegraphics[width=0.7\columnwidth]{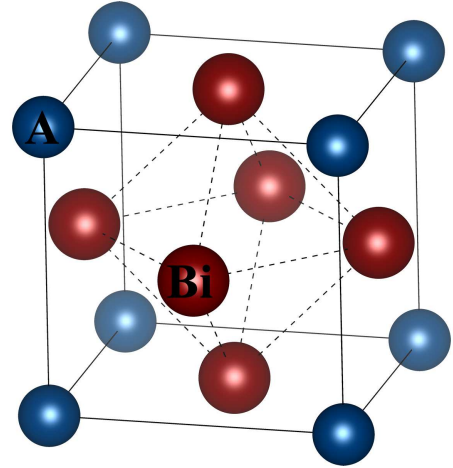}
\caption{ Crystal structure of ABi$_{3}$.}
\label{Fig_1_structure}
\end{figure}

There is a class of Bi-rich superconductors ABi$_{3}$ (A=Sr and Ba) with
simple AuCu$_{3}$ structure (Fig. \ref{Fig_1_structure}). Polycrystalline
ABi$_{3}$ (A=Sr and Ba) and the superconductivity were firstly reported
by Matthias and Hulm in 1952 \cite{Matthias-ABi3}.
Subsequently, to the best of our knowledge, there were only one experimental
report about the polycrystalline samples of Eu doped SrBi$_{3}$ in the
following 60 years \cite{Kempf-Eu-SrBi3}. First principle calculation
without including SOC estimated a superconducting transition temperature
(Tc) of 1.8 K for SrBi$_{3}$ \cite{SrBi3-band}, which is remarkably smaller
than the experimentally measured $T_{c}$ of $\sim$5.6 K \cite{Matthias-ABi3,
Kempf-Eu-SrBi3}. Such large deviation was attributed to the disadvantage
of the calculation method \cite{SrBi3-band}. Few people have realized
that SOC should influence the superconductivity of those compounds
in the past years. Very recently, ABi$_{3}$ (A=Sr and Ba) were reinvestigated
\cite{BaBi3-cava, Iyo-SrBi3-Na}. Haldolaarachchige et al. prepared
the single crystal sample of BaBi$_{3}$ and concluded the physical parameters
in detail \cite{BaBi3-cava}. Iyo et al. investigated superconductivity
in polycrystalline sample of Na doped SrBi$_{3}$ \cite{Iyo-SrBi3-Na}. However,
the role of SOC still has not been discussed. 

In this work, we figured out the role of SOC in ABi$_{3}$ (A=Sr and Bi)
by theoretical investigation of the band structures, phonon properties,
and electron-phonon coupling. We found that without including SOC,
strong Fermi surface nesting exists between the electron-pockets at
the face centers, which leads to phonon instabilities. SOC suppresses
the nesting and stabilize the phonon modes. Moreover, we found the
calculation without including SOC largely underestimates $T_{c}$, while
with SOC the calculated $T_{c}$ are very close to those determined in experiments
performed using single crystal samples. Our investigation demonstrates
that superconductivity in Bi-rich compounds ABi$_{3}$ (A=Sr and Bi) is
strongly enhanced by SOC, which is due to not only the SOC induced
softening, but also the SOC related increase of electron-phonon coupling
matrix elements. Furthermore, the Bi atoms in the (111) plane of ABi$_{3}$
(A=Sr and Bi) is trigonal, which is very similar to situations in
the Bi plane of Bi$_{2}$Se$_{3}$ and ultrathin Bi (111) Films. Therefore, the
Bi-rich superconductor ABi$_{3}$ (A=Sr and Bi) can be a potential platform
to construct a heterostructure of superconductor/topological insulator
to realize topological superconductivity.

\section{Methods}
The density functional theory (DFT) calculations were carried out using QUANTUM ESPRESSO package \cite{QE} with ultrasoft pseudopotentials. The exchange-correlation interaction was treated with the generalized gradient approximation (GGA) with Perdew-Burke-Ernzerh (PBE) of parametrization \cite{PBE}. The energy cutoff for the plane-wave basis set was 40 Ry. Brillouin zone sampling is performed on the Monkhorst-Pack (MP) mesh \cite{MP} of $16 \times 16\times 16$, while a denser  $32\times32\times32$  grid was used in the electron phonon coupling calculations. The Vanderbilt-Marzari Fermi smearing method with a smearing parameter of $\sigma=0.02$ Ry was used for the calculations of the total energy and electron charge density. Phonon dispersions were calculated using density functional perturbation theory (DFPT) with a $4\times4\times4$ mesh of $q$-points. To investigate the effect of spin-orbit coupling, fully relativistic calculations were carried out.

Single crystalline specimens of SrBi$_{3}$ were prepared by Bi-self flux. Sr (99.9\%, Alfa Aaser) and  Bi (99.99\%, Alfa Aaser) with mole ratio 1:6 were loaded into alumina crucible, which was placed in quartz tube inside an Ar-filled box. The quartz tubes were sealed under a vacuum. The sealed quartz tubes were slowly heated to 600 $^{\circ}$C for 10 hours, then slowly cooling to 330 $^{\circ}$C with 3 $^{\circ}$C/h. Finally, the excess Bi-flux was removed by decanting. Rectangular shape single crystals with shining surface were observed. The size is about $3\times3\times2$ mm$^{3}$. The single crystals were kept inside the glove box until characterization. Such handling is necessary to avoid decomposition.  Powder X-ray diffraction (XRD) patterns were taken with Cu $K_{\alpha1}$ radiation ($\lambda=0.15406$ nm) using a PANalytical X’pert diffractometer at room temperature. Magnetic, electrical transport and heat capacity measurements  were measured using the Quantum Design MPMS-XL5 and PPMS-9. Magnetization measurements under pressure were performed using a pistoncylinder apparatus using the gasket and glycerol as the pressure transmitting medium.

\section{Results and discussions}
The structures of ABi$_{3}$ (A=Sr and Bi) were fully optimized with respect to lattice parameter and atomic positions. For SrBi$_{3}$, the optimized lattice parameter is 5.055 \AA, which is in good agreement with experimental value \cite{JPCS-1970}. Nonmagnetic (NM), ferromagnetic (FM), and antiferromagnetic (AFM) states are tested in the system. The magnetic moments of each atom in FM and AFM states are converged to zero, which is consistent with the NM ground state measured in experiment. 

\begin{figure}
\includegraphics[width=0.99\columnwidth]{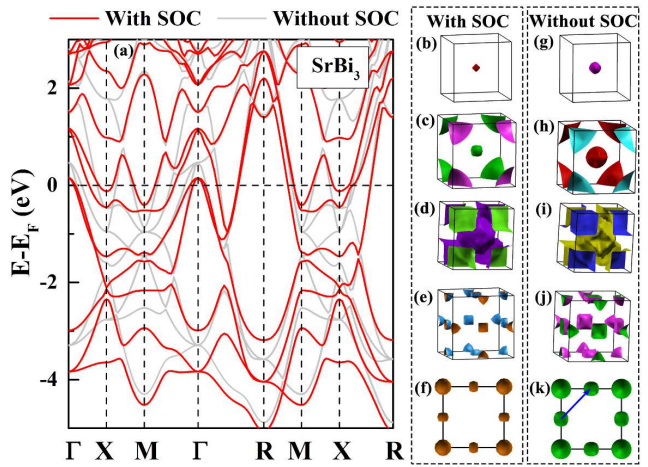}
\caption{ (a) The band dispersion of SrBi$_{3}$ with (the red lines) and without (the grey lines) SOC. (b)-(e) are the Fermi surface of SrBi$_{3}$ with SOC, while (g)-(j) are those without SOC. (f) and (k) are the middle cross sections of (e) and (j). The blue arrow in (k) denotes the nesting vector $M$. }
\label{Fig_2_SrBi3_band_FS}
\end{figure}

In Fig. \ref{Fig_2_SrBi3_band_FS} (a) we compared the band dispersion of SrBi$_{3}$ with and without including SOC. Because of the high concentration of Bi, one can note that SOC remarkably lifts band degeneracy near Fermi energy ($E_{F}$) in all the symmetry directions. Four bands cross $E_{F}$ in each case. SOC shrinks the volumes and marginally changes the shapes of the Fermi surfaces in SrBi$_{3}$, while the locations of the Fermi surfaces are unchanged. More specifically, there are five hole pockets and two electron pockets. Three hole pockets locate around $\Gamma$ and the rest two hole pockets locate around $R$ (Figs. \ref{Fig_2_SrBi3_band_FS} (b)-(d) and (g)-(i)). Two electron pockets locate around $M$ and $X$ points, respectively (Figs. \ref{Fig_2_SrBi3_band_FS} (e) and (j)). 

\begin{figure}
\includegraphics[width=0.99\columnwidth]{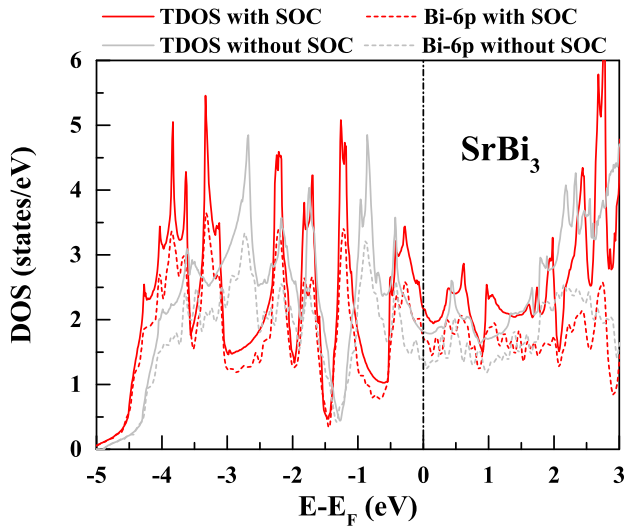}
\caption{ The DOS of SrBi$_{3}$ with (red) and without (grey) SOC. The solid and dashed lines denote the TDOS and the contribution of $6p$ electrons of Bi, respectively. }
\label{Fig_3_SrBi3_DOS}
\end{figure}

The density of states (DOS) of SrBi$_{3}$ with and that without SOC were also compared. As shown in Fig. \ref{Fig_3_SrBi3_DOS}, one can note the total DOS (TDOS) near $E_{F}$ are predominately contributed by Bi-$6p$ electrons (Fig. \ref{Fig_3_SrBi3_DOS}). SOC increases the TDOS at $E_{F}$ ($N(E_{F})$)  by $\sim20\%$ (Table \ref{Table_1}).

\begin{table}

\caption{The calculated $N(E_{F})$, $\omega_{log}$, $\lambda$, and $T_{c}$
of ABi$_{3}$ (A=Sr and Bi) with and without SOC. For BaBi$_{3}$ without SOC,
since the phonon modes are unstable, $\omega_{log}$ , $\lambda$,
and $T_{c}$ were not calculated. }

\begin{tabular}{ccccc}
\hline 
 & $N(E_{F})$ (states/eV) & $\omega_{log}$ (K) & $\lambda$ & $T_{c}$ (K)\tabularnewline
\hline 
SrBi$_{3}$ without SOC & 1.81 & 63.03 & 0.91 & 3.73\tabularnewline
SrBi$_{3}$ with SOC & 2.17 & 64.04 & 1.11 & 5.15\tabularnewline
BaBi$_{3}$ without SOC & 2.02 & -- & -- & --\tabularnewline
BaBi$_{3}$ with SOC & 2.40 & 48.69 & 1.43 & 5.29\tabularnewline
\hline 
\end{tabular}\label{Table_1}
\end{table}

\begin{figure}
\includegraphics[width=0.99\columnwidth]{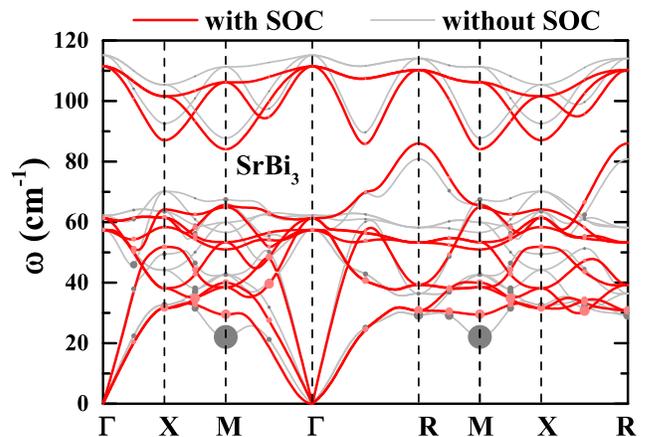}
\caption{
The phonon dispersions of of SrBi$_{3}$ with (red) and without (grey) SOC.
The phonon dispersions are decorated with symbols, proportional to
the partial electron-phonon coupling strength $\lambda_{\mathbf{q}}^{\nu}$. }
\label{Fig_4_SrBi3_phonon}
\end{figure}

Figure \ref{Fig_4_SrBi3_phonon} shows the phonon dispersions of SrBi$_{3}$. In most directions, SOC softens the phonon modes. However, one can note a remarkable softening in the lowest acoustic mode at M point appears when SOC is not included. We attribute such instability to the Fermi surface nesting between the electron pockets around the face centers ($X$ point) of the Brillouin zone. As shown in Figs. \ref{Fig_2_SrBi3_band_FS} (j) and (k), when SOC is not included, the electron pockets at face centers in SrBi$_{3}$ show the swelling cubic shape. Large fragments of the pockets at different face centers can be coupled by the nesting vector $M$ (Fig. \ref{Fig_2_SrBi3_band_FS} (k)). Therefore, stronger instability at M was shown in SrBi$_{3}$ without SOC. On the other hand, SOC changes the shape of such pockets into rectangular hexahedron (Figs. \ref{Fig_2_SrBi3_band_FS} (e) and (f)), which suppresses the nesting and stabilizes the phonon mode at $M$.

\begin{figure}
\includegraphics[width=0.99\columnwidth]{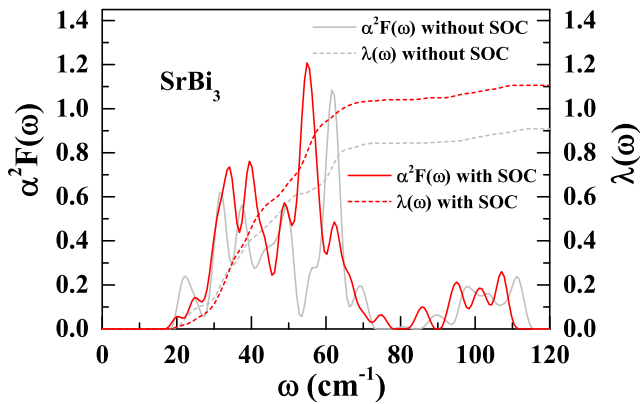}
\caption{Eliashberg function (left) and the integrated electron-phonon coupling strength (right) for SrBi$_{3}$ with (red) and without (grey) SOC, respectively. }
\label{Fig_5_SrBi3_a2f}
\end{figure}

The electron-phonon coupling can be qualitatively discussed based on Hopfield expression:\begin{equation}
\lambda=\frac{N(E_{F})D^{2}}{M\omega^{2}},
\end{equation} where $D$ is the deformation potential, and $M$ is the atomic mass. In SrBi$_{3}$, SOC largely increases $N(E_{F})$ and softens most phonon modes. Therefore, one can expect a stronger eletron-phonon coupling when SOC is included. More specifically, Fig. \ref{Fig_5_SrBi3_a2f} shows the Eliashberg spectral function:
\begin{equation}
\alpha^{2}F(\omega)=\frac{1}{N(E_{F})}\underset{\mathbf{k},\mathbf{q},\nu,n,m}{\sum}\delta(\epsilon_{\mathbf{k}}^{n})\delta(\epsilon_{\mathbf{k}+\mathbf{q}}^{m})\mid g_{\mathbf{k},\mathbf{k}+\mathbf{q}}^{\nu,n,m}\mid^{2}\delta(\omega-\omega_{\mathbf{q}}^{\nu}),\label{eq:a2f}
\end{equation} where $\omega_{\mathbf{q}}^{\nu}$ is the phonon frequency, $\epsilon_{\mathbf{k}}^{n}$ is the electronic energy, and  $g_{\mathbf{k},\mathbf{k}+\mathbf{q}}^{\nu,n,m}$ is the electron-phonon coupling matrix element. The total electron-phonon coupling strength is \begin{equation}
\lambda=2\int_{0}^{\infty}\frac{\alpha^{2}F(\omega)}{\omega}d\omega=\underset{\mathbf{q\nu}}{\sum}\lambda_{\mathbf{q}}^{\nu},\label{eq:lambda}
\end{equation} where the electron-phonon coupling strength for each mode ($\lambda_{\mathbf{q}}^{\nu}$) is defined as,
\begin{equation}
\lambda_{\mathbf{q}}^{\nu}=\frac{\gamma_{\mathbf{q}}^{\nu}}{\pi\hbar N(E_{F})M\omega_{\mathbf{q}}^{\nu}}
\end{equation} which are visualized as circles in Fig. \ref{Fig_4_SrBi3_phonon}. According to this definition, phonon modes with a lower frequency will lead to stronger electron-phonon coupling. When SOC is not included, the large softening of lowest acoustic mode around $M$ point contributes a stronger electron-phonon coupling compared with the case for that SOC is included (Fig. \ref{Fig_4_SrBi3_phonon}.). However, it only leads to a small peak between 20 to 25 cm$^{-1}$, which contributes only $\sim10\%$ of the total electron-phonon coupling strength (Fig. \ref{Fig_5_SrBi3_a2f}). For the modes between 30 to 40 cm$^{-1}$, the $\alpha^{2}F(\omega)$ peaks with SOC are notably higher than those when SOC is not included, indicating SOC has a sizable enhancement in the electron-phonon coupling matrix elements. Furthermore, since SOC softens the modes in most directions, above 40 cm$^{-1}$ the peaks with SOC become stronger and have lower frequencies. As shown in Fig.  \ref{Fig_5_SrBi3_a2f}, SOC largely  increased ($\sim20\%$) the total electron-phonon coupling strength (Table \ref{Table_1}).

We estimated Tc based on the Allen-Dynes formula \cite{Allen-Dynes}:
\begin{equation}
T_{C}=\frac{\omega_{log}}{1.2}\exp\left(-\frac{1.04(1+\lambda)}{\lambda-\mu^{*}-0.62\lambda\mu^{*}}\right)\text{,}\label{eq:TC}
\end{equation}
 where  the Coulomb pseudopotential $\mu^{*}$ is set to a typical value of $\mu^{*}=0.1$.   The logarithmically averaged characteristic phonon frequency $\omega_{log}$
is defined as
\begin{equation}
\omega_{log}=\exp\left(\frac{2}{\lambda}\int\frac{d\omega}{\omega}\alpha^{2}F(\omega)\log\omega\right).
\end{equation} We listed the calculated $T_{C}$ and $\omega_{log}$ in Table \ref{Table_1}. When SOC is not included, the calculated  $T_{C}$  is only 3.73 K. While, with inclusion of SOC, the calculated Tc is 5.15 K. It indicates that the importance of SOC in the superconductivity of SrBi$_{3}$.

\begin{figure}
\includegraphics[width=0.99\columnwidth]{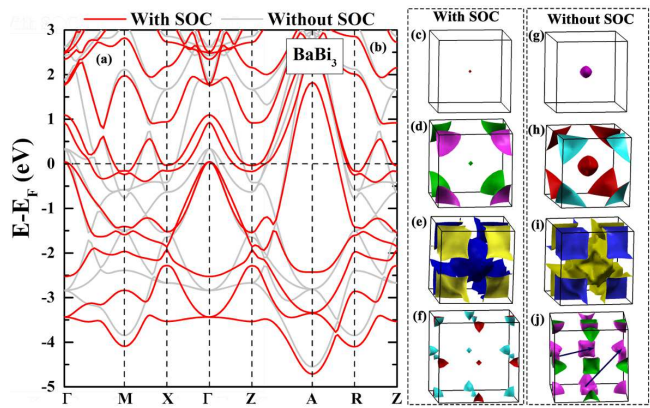}
\caption{(a) The band dispersion  of BaBi$_{3}$ with (the red lines) and without (grey lines) SOC.  (b)-(e) are the Fermi surface of BaBi$_{3}$ with SOC, while (f)-(i) are those without SOC. The blue arrows in (i) denotes the nesting vectors $M$ and $R$. }
\label{Fig_6_BaBi3_band_FS}
\end{figure}

We also calculated the properties of BaBi$_{3}$. The substitution of Sr by Ba changes the crystal from cubic to tetragonal structure. However, the lattice parameters of $a$ (5.188 \AA) and that of $c$ (5.136 \AA) are very close to each other. Therefore, the resulted band structure and Fermi surface of BaBi$_{3}$ (Fig. \ref{Fig_6_BaBi3_band_FS}) are very similar to those of SrBi$_{3}$. Our calculation is in good agreement with previous report \cite{BaBi3-cava}. SOC remarkably lifts the band degeneracy near Fermi energy ($E_{F}$) in all the symmetry directions of BaBi$_{3}$  as well (Fig. \ref{Fig_6_BaBi3_band_FS} (a)). Four bands cross $E_{F}$, formatting three hole pockets around the body center of the Brillouin Zone ($\Gamma$), two hole pockets around the corner of the Brillouin Zone ($A$), and two electron pockets locating at the face centers (
$X$ and $Z$) and edge centers ($M$ and $R$), respectively (Figs. \ref{Fig_6_BaBi3_band_FS} (b)-(i)). 

\begin{figure}
\includegraphics[width=0.99\columnwidth]{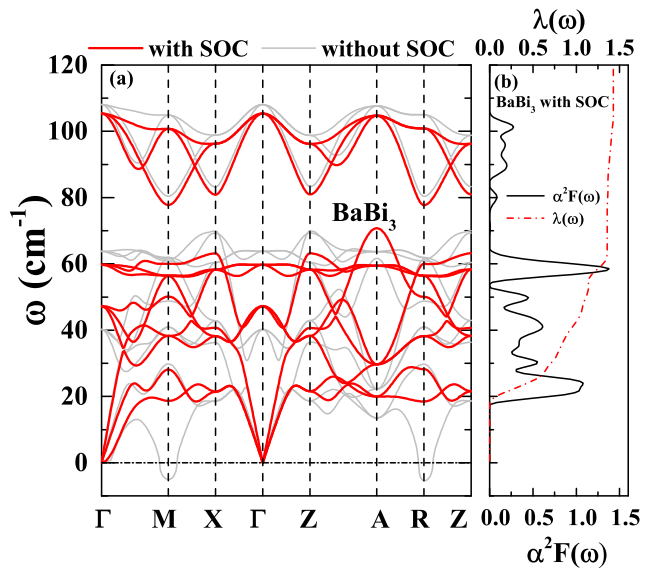}
\caption{(a) The phonon dispersions of of BaBi$_{3}$ with (red) and without (grey) SOC. (b) Eliashberg function (bottom) and the integrated electron-phonon coupling strength (top) for BaBi$_{3}$ with SOC.}
\label{Fig_7_BaBi3_phonon}
\end{figure}

Figure \ref{Fig_7_BaBi3_phonon} (a) shows the phonon dispersion of BaBi$_{3}$. Similar to SrBi$_{3}$, when SOC is not included, the nesting between the electron pockets at different face centers leads to very strong instabilities with imaginary frequency at $M$ and $R$. SOC changes such swelling cubic-like electron pockets into spindle-shaped pockets. Therefore, the instabilities are suppressed. In other words, SOC stabilizes the structure of BaBi$_{3}$. The calculated Eliashberg function of BaBi$_{3}$ with SOC is shown in Fig.  \ref{Fig_7_BaBi3_phonon} (b). The calculated total electron-phonon coupling strength is 1.43, leading to $T_{c}$ of 5.29 K.

A convenient way to prove our calculation is directly comparing the calculated $T_{c}$ with the experimentally obtained ones.  Although SrBi$_{3}$ has been synthesized sixty years ago, the reported data are mainly based on the SrBi$_{3}$ polycrystalline samples  \cite{Matthias-ABi3,Kempf-Eu-SrBi3} and the comprehensive studied on SrBi$_{3}$ single crystal is rarely reported. As we know, the superconductivity is very sensitive to the sample quality of polycrystalline. For example, the reported $T_{c}$ of MgCNi$_{3}$ in polycrystalline samples varies from 6 K to 9 K \cite{MgCNi3-review}. On the other hand, single crystal with good sample quality can reflect the intrinsic properties of the material. The $T_{c}$ of MgCNi$_{3}$ is proved to be $\sim$6.7 K using single crystal sample, while the physical parameters are measured with higher accuracy in single crystal as well. For the present Bi-rich compounds ABi$_{3}$ (A=Sr and Ba), the studies on single crystal samples are necessary to prove our estimation. Previously Haldolaarachchige et al. \cite{BaBi3-cava} prepared the single crystal of BaBi$_{3}$ and measured the physical properties. Our calculated $T_{c}$ of 5.29 K is very close to the measured $T_{c}$ of 5.95 K. Here we synthesized the single crystal of SrBi$_{3}$ and performed the related physical measurements. 

\begin{figure}
\includegraphics[width=0.99\columnwidth]{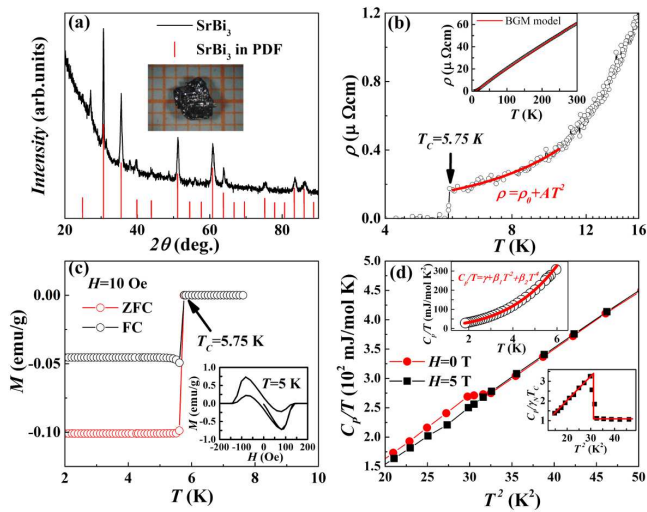}
\caption{(a) Powder XRD pattern of SrBi$_{3}$ crushed from many single crystals. The red bars are SrBi$_{3}$ in PDF card. The inset shows the studied SrBi$_{3}$ single crystal. (b) Temperature dependence of resistivity of the polished SrBi$_{3}$ single crystal. The solid line is the Fermi liquid fitting at the low temperature. The inset shows the Bloch-Grüneisen-Mott (BGM) model fitting of the resistivity. (c) ZFC and FC magnetic susceptibility of SrBi$_{3}$ single crystal measured at $H=10$ Oe. The superconducting temperature $T_{c}$ is 5.75 K. The inset shows the magnetic field dependence of magnetization at $T=5$ K. (d) Heat capacity of SrBi$_{3}$ single crystal measured under $H=0$ T and $H=5$ T. The upper inset shows the $\frac{C_{p}}{T}$ versus $T$, the solid line is fitting according to $\frac{C_{p}}{T}=\gamma+\beta_{1}T^{2}+\beta_{2}T^{4}$. The lower inset shows the $\frac{C_{p}}{\gamma_{N}T}$ versus $T^2$. }
\label{Fig_8_SrBi3_rt}
\end{figure}
As shown in Fig. \ref{Fig_8_SrBi3_rt} (a), single crystals  with a size of $3\times3\times2$ mm$^{3}$   were obtained. Powder XRD measurement indicates the good sample quality. The measured temperature dependences of the resistivity ($\rho$), magnetization ($M$), and specific heat ($C_{p}$) show the superconducting transition at 5.75 K, which is very close to our estimation. Moreover, the electronic specific heatγ, which is obtained from the fitting of specific heat based on the relation ( $\frac{C_{p}}{T}=\gamma+\beta_{1}T^{2}+\beta_{2}T^{4}$, shows a value of 10.249 mJ/mol K$^2$. From the relation $\gamma=\frac{\pi^{2}k_{B}^{2}}{3}N(E_{F})(1+\lambda)$, using the calculated $N(E_{F})$=2.17 states/eV, we can estimated the electron-phonon coupling parameter $\lambda=1.005$, which is very close to our calculated $\lambda=1.11$. The ratio $\frac{\Delta C}{\gamma T_{c}}=2.12$ is higher than the BCS weak-coupling limit of 1.43, which also supports our estimated strong coupling scenario. Other fitted physical parameters are presented in the supplemented file. All the measurements verifies our calculation.

\section{Conclusion}

In this work, we figured out the role of SOC in ABi$_{3}$ (A=Sr and Bi) by theoretical investigation of the band structures, phonon properties, and electron-phonon coupling. We found that when SOC is not included, strong Fermi surface nesting exists between the electron-pockets at the face centers, which leads to phonon instability. SOC suppresses the nesting and stabilize the phonon modes. Moreover, we found the calculation without including SOC largely underestimates $T_{c}$. With SOC, the calculated Tc are very close to the $T_{c}$ determined in measurements on single crystal samples. Our investigation demonstrates that superconductivity in Bi-rich compounds ABi$_{3}$ (A=Sr and Bi) is strongly enhanced by SOC, which is due to not only the SOC induced softening, but also the SOC related increase of electron-phonon coupling matrix elements. Since the arrangement of Bi atoms in the (111) plane of ABi$_{3}$ (A=Sr and Bi) is very similar to that in the Bi plane of Bi$_{2}$Se$_{3}$ and that in ultrathin Bi(111) Films, the Bi-rich superconductor ABi$_{3}$ (A=Sr and Bi) can be a potential platform to construct a heterostructure of superconductor/topological insulator to realize topological superconductivity.

\begin{acknowledgments}
This work was supported by the Joint Funds of the National Natural Science Foundation of China and the Chinese Academy of Sciences' Large-Scale Scientific Facility under contracts (U1432139, U1232139), the National Nature Science Foundation of China under contracts (11304320, 11274311, 51171177, 11474289), the National Key Basic Research under contract 2011CBA00111, and the National Nature Science Foundation of Anhui Province under contract 1508085ME103. 
\end{acknowledgments}

\end{document}